\documentclass[reprint,amsmath,amssymb,aps]{revtex4-2}
\usepackage{soul}
\usepackage{graphicx}
\usepackage{dcolumn}
\usepackage{bm}
\usepackage{xcolor,cancel}
\usepackage{xcolor}

\begin{document}

\preprint{APS/123-QED}

\title{Deformed neutron stars}

\author{J.T. Quartuccio}
 \email{jon.quartuccio@gmail.com}
\affiliation{Laborat\'orio de F\'isica Te\'orica e Computacional (LFTC),
 Universidade Cidade de S\~ao Paulo (UNICID) - Rua Galv\~ao Bueno 868, 01506-000 S\~ao Paulo, Brazil}
\author{P.H.R.S. Moraes}
 \email{moraes.phrs@gmail.com}
\affiliation{Laborat\'orio de F\'isica Te\'orica e Computacional (LFTC),
 Universidade Cidade de S\~ao Paulo (UNICID) - Rua Galv\~ao Bueno 868, 01506-000 S\~ao Paulo, Brazil}
\author{J.D.V. Arba\~nil}
\email{jose.arbanil@upn.pe}
\affiliation{Departamento de Ciencias, Universidad Privada del Norte, Avenida el Sol 461 San Juan de Lurigancho, 15434 Lima, Peru}
\affiliation{Facultad de Ciencias F\'isicas, Universidad Nacional Mayor de San Marcos, Avenida Venezuela s/n Cercado de Lima, 15081 Lima, Peru}

\begin{abstract}
\begin{center}
\end{center}
We present solutions for non-spherically symmetric neutron stars. We begin by deriving the Tolman-Oppenheimer-Volkoff equations from a parameterized metric that takes into account the deformation of the star due to differences in equatorial and polar pressures, expressed in terms of a parameter $\mathcal{D}$, which is the ratio between polar and equatorial radius. The stellar structure is solved using the GM1 equation of state and the Tolman-Oppenheimer-Volkoff equations for deformed objects are numerically integrated using the fourth-order Runge-Kutta method for different values of the parameter $\mathcal{D}$. We show that larger values of $\mathcal{D}>1$, that describe prolate neutron stars, yield smaller values of mass and radius, while for smaller values of $\mathcal{D}<1$, describing oblate neutron stars, larger values for mass and radius are attained. From the confrontation of our model theoretical predictions with recent observational data on pulsars, it is possible to constrain the values of the parameter $\mathcal{D}$.\\


\end{abstract}  

\maketitle

\section{Introduction}\label{sec:int}

Compact astrophysical objects, such as neutron stars (NSs), have been the main subject of study of several researchers for some time. The LIGO's (Laser Interferometer Gravitational wave Observatory) first detection of gravitational waves (GW150914) \cite{abbott/2016}, coming from a binary system of black holes, strongly contributed to that. Later, the first detection of a binary NS system occurred (GW170817) \cite{abbott/2017}, which together with the electromagnetic counterpart detection of the event \cite{abbott/2017b}, gave birth to the so-called multi-messenger era of Astrophysics. 

The NICER (Neutron star Interior Composition ExploreR) \cite{gendreau/2017} is a NASA mission dedicated to the study of NSs. The radii of PSR J0740+6620 and PSR J0030+0451 were estimated from this experiment, respectively, in \cite{salmi/2022} and \cite{riley/2019}. The possibility of estimating the NS radius is of unprecedented importance for NS astrophysics because we still do not know the NS equation of state (EoS). Several E'soS have been proposed for the NS interior, among which we quote  \cite{baym/2019,somasundaram/2023,pratten/2022}. 

Even before the first gravitational wave detection, gravitational wave astrophysics was proposed as a tool to constraint the NS EoS \cite{read/2009,del_pozzo/2013}, which has come true as one can see, for instance, the constraints obtained in light of GW170817 \cite{radice/2018,raithel/2019,zhu/2018}.

In 2020, an outstanding compact astrophysical object was reported as a component of a coalescing binary system whose gravitational wave sign was detected (GW190814) \cite{abbott/2020}. GW190814 progenitor system is known as a $22.2-24.3$M$_\odot$ black hole and a $2.50-2.67$M$_\odot$ {\it compact object}, namely a black hole or a NS. The secondary object is, remarkably, either the lightest black hole or the heaviest NS ever discovered in a binary system. The majority of studies have been pointing to the secondary object in GW190814 as a NS \cite{huang/2020,wu/2021,godzieba/2021,tan/2020,tsokaros/2020}, although the possibility of a black hole cannot be discarded, as one can check \cite{vattis/2020}.   

Such a massive NS is not straightforwardly modeled. In \cite{horvath/2021}, for instance, the authors have considered anisotropic quark matter to simultaneously describe PSR J0030+0451 and the secondary component in GW190814 as a NS. Even extended gravity formalisms have been used to explain the NS in GW190814, as one can check  \cite{astashenok/2020,astashenok/2021}, for instance.

In the case of matter fields at high densities ($\rho>10^{15}$g/cm$^3$), anisotropy may arise naturally \cite{ruderman/1972,canuto/1974} and play a fundamental role in the interior of compact astrophysical objects. Electromagnetic and fermionic fields in NSs \cite{sawyer/1973} and superfluidity \cite{carter/1998} are natural examples of anisotropy. 

There are plenty of works devoted to the study of anisotropic spherically symmetric static stellar configurations, as one can check References \cite{mak/2003,mak/2002,harko/2002,errehymy/2021,folomeev/2015}. However, anisotropy can imply in deviations from spherical symmetry, namely deformed compact astrophysical objects \cite{zubairi/2017,zubairi/2015}.

In the present article, we will construct static equilibrium configurations of deformed NSs ($\mathcal{D}$NSs). The article is presented as follows. In Section \ref{sec:dtov}, starting from the metric that evades spherical symmetry, we are going to carefully derive the deformed version of the Tolman-Oppenheimer-Volkoff (TOV) equation \cite{tolman/1939,oppenheimer/1939}, which we shall refer to as ``$\mathcal{D}-$TOV equations''. In Section \ref{sec:sol}, we are going to present the EoS we will use to describe matter inside $\mathcal{D}$NSs, the boundary conditions we assume, our numerical method and the solutions. We also confront our solutions with observational data. Lastly, we discuss our results and present our concluding remarks in Section \ref{sec:dc}. 

\section{Deformed object metric and energy-momentum tensor, and the derivation of the $\mathcal{D}-$TOV equations}\label{sec:dtov}

In the present section, we show how to obtain the TOV equations for a deformed object, which we will refer to as the $\mathcal{D}-$TOV equations. We can analyze deformation by evoking the following parameterized metric \cite{zubairi/2017}

\begin{equation}\label{dtov1}
ds^2 = -e^{2\phi(r)}dt^2 + \Big[ 1 - \dfrac{2m(r)}{r}\Big]^{-\mathcal{D}}dr^2+r^2(d\theta^2 + \sin^2 \theta d\phi^2).
\end{equation}
In the metric presented in Equation (\ref{dtov1}) above, in which we have assumed the speed of light $c$ to be taken as $1$, as it will be done throughout the article, $\phi(r)$ is a metric potential and $m(r)$ is the gravitational mass of the studied object. The deformation parameter $\mathcal{D}$ provides the ratio between the polar and equatorial radius, i.e., $\mathcal{D}\equiv z/r$. Spherical symmetry is recovered when $\mathcal{D}=1$. In Figure 1 below we exemplify how the deformation parameter works. 

\begin{figure}[h]
    \centering
    \includegraphics[scale=0.5]{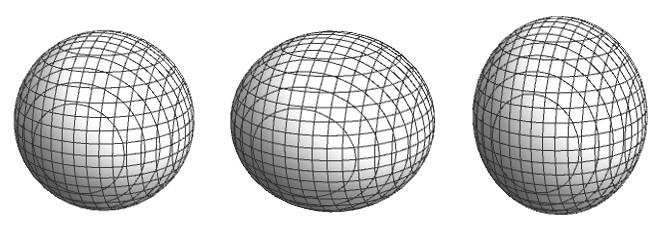}
    \caption{Different geometries depending on the value of $\mathcal{D}$. In the image on the left there is no deformation, therefore, in this case, the star has spherical symmetry and $\mathcal{D}=1$. In the central image, $\mathcal{D}<1$, and we have an oblate spheroid, while in the image on the right, $\mathcal{D}>1$, and we have a prolate spheroid.}
    \label{fig:enter-label}
\end{figure}


The energy-momentum tensor 

\begin{equation}\label{tmomentum}
	T_{\mu \nu} =
	\begin{pmatrix}
		\rho&0&0&0\\
		0&p_{r}&p_{rz}&0\\
		0&p_{rz}&p_{z}&0\\
		0&0&0&p_{rz}
	\end{pmatrix}
\end{equation}
describes the deformed compact object \cite{zubairi/2017}, for which $\rho$ represents the energy density, $p_{r}$ is the pressure that acts in the radial direction, while $p_{z}$ is the pressure in polar direction and $p_{rz}$ is the pressure in $r-z$ direction.

From the $rr$ component of the Einstein's field equations of General Theory of Relativity for the above metric and energy-momentum tensor, we obtain



\begin{equation}\label{phi}
	\phi'(r) = \dfrac{4\pi r^2 p_r-\dfrac{1}{2}\Big(1 - \dfrac{2m}{r}\Big)^\mathcal{D} + \dfrac{1}{2}}{\Big(1 - \dfrac{2m}{r}\Big)^\mathcal{D}r}. 
\end{equation}
Such an equation, when equalized with the covariant derivative of the energy-momentum tensor, yields:

\begin{equation}\label{dtovr}
    \dfrac{dp_r}{dr} =- \dfrac{(\rho + p_r) \Big[8\pi r^2 p_r -\Big(1 - \dfrac{2m}{r}\Big)^\mathcal{D} + 1\Big]}{2\Big(1 - \dfrac{2m}{r}\Big)^{\mathcal{D}}r}.
\end{equation}

If we use the above definition of the deformation parameter $\mathcal{D}$ in Eq.\eqref{dtovr}, we can rewrite this equation as

\begin{equation}\label{dtovz}
   \dfrac{dp_z}{dz} = - \dfrac{\mathcal{D}(\rho + p_z) \Big[8\pi \Big(\dfrac{z}{\mathcal{D}}\Big)^2 p_z - \Big(1 - \dfrac{2m\mathcal{D}}{z}\Big)^\mathcal{D} + 1\Big]}{2\Big(1 - \dfrac{2m\mathcal{D}}{z}\Big)^{\mathcal{D}}z}.
\end{equation}

We now have two $\mathcal{D}$-TOV equations: the first (\ref{dtovr}) works with the parallel pressure, associated with $r$, and the second (\ref{dtovz}), with the perpendicular pressure, associated with $z$. If the star is not deformed, the parameter $\mathcal{D}$ is equal to 1 and we retrieve the usual TOV equation.

According to \cite{zubairi/2015}, the above gravitational mass $m$ of the system in terms of $r$ and $z$ is given as follows

\begin{equation}\label{eq_mass}
    \dfrac{dm}{dr} = 4\mathcal{D}\rho \pi r^2.
\end{equation}


\section{Results}\label{sec:sol}

\subsection{Equation of state and anisotropic profile}\label{ss:esos}

We employ the mean-field model to analyze the stellar structure configurations. We use the standard lagrangian, which describes matter made up of nucleons, hyperons and electrons, of the form \cite{lugones2010}:
\begin{eqnarray}\label{EoS}
&&\mathcal{L}_H=\sum_{B}{\bar \psi}_B\bigg[\gamma_\mu\left(i\partial^\mu-g_{\omega B}\omega^\mu-\frac{1}{2}g_{\rho B}\vec{\tau} \cdot \vec{\rho}^{\,\mu}\right)\nonumber\\
&&-\left(m_B-g_{\sigma B}\sigma\right)\bigg]\psi_B+\frac{1}{2}\left(\partial_\mu \sigma\partial^{\mu} \sigma-m_\sigma^2\sigma^2\right)\nonumber\\
&&-\frac{1}{3}bm_n(g_\sigma \sigma)^3-\frac{1}{4}c(g_\sigma \sigma)^4-\frac{1}{4}\omega_{\mu\nu} \omega^{\mu\nu}\nonumber\\
&&+\frac{1}{2}m_\omega^2 \omega_{\mu} \omega^{\mu}-\frac{1}{4}\vec{\rho}_{\mu\nu} \cdot \vec{\rho}^{\,\mu\nu}+\frac{1}{2}m_\rho^2\vec{\rho}_{\mu} \cdot \vec{\rho}^{\,\mu}\nonumber\\
&&+\sum_L{\bar \psi}_L\left[i\gamma_\mu\partial^\mu-m_L\right]\psi_L.
\end{eqnarray}
In (\ref{EoS}), $L$ refers to leptons and they are treated like non-interacting, $B$ represents baryons coupled to the scalar meson $\sigma$, $\omega_\mu$ depicts the isoscalar-vector meson and $\rho_\mu$ is the isovector-vector meson. ${\bar\psi}_B$ is the conjugate adjoint of the fermionic field $\psi_B$ associated with $B$ while $\gamma_\mu$ are the Dirac matrices. The isospin matrices are represented by $\vec{\tau}$, $m_B$ is the mass of the baryon, while $m_\sigma$, $m_\omega$ and $m_\rho$ are the masses associated with $\sigma$, $\omega$ and $\rho$. The mass of the nucleon is given by $m_n$, while $m_L$ is the mass of the lepton. The coupling constants are given by $g_{\omega B}, g_{\sigma B}$ and $g_{\rho B}$. The ${\bar \psi}_L$ term represents the adjoint conjugate of the fermionic field $\psi_L$, $b$ and $c$ are constants that characterize the strength of interactions, ${\bar \rho}_{\mu \nu}$ represents the field strength tensor associated with the vector meson field ${\bar \rho}_\mu$. The five constants are fitted to the bulk properties of nuclear matter; in this case, the GM1 parametrization is employed \cite{glendenning1991}. The EoS $p_r(\rho)$ derived from \eqref{EoS} and the coupling constants taken into account for the GM1 parametrization are shown in \cite{lugones2010}. At low densities, the Baym, Pethick, and Sutherland \cite{BPS1971} model is used.

For the anisotropic profile, we use 

\begin{equation}
p_r=p_z+(\mathcal{D}-1)\sigma.
\end{equation}
When $\mathcal{D}=1$, the isotropic case is recovered. This relation allows us to have regularity at the star center. 

\subsection{Numerical method}\label{ss:bcnm}

To study the anisotropic influence in the stellar equilibrium configuration of compact stars - once defined the EoS and the anisotropic profile - the structure equations \eqref{dtovr}, \eqref{dtovz} and \eqref{eq_mass} are numerically integrated from the center ($r=0$) to the surface of the star $r=R$, by using the Runge-Kutta fourth order method, for a given central energy density $\rho_c$ and $\mathcal{D}$.

The solution begins with the values at $r=0$, namely:
\begin{equation}
\rho(0)=\rho_c,\quad\sigma(0)=0,\quad m(0)=0,
\end{equation}
and ends when the star surface is found ($p_r(R)=0$).

\subsection{Equilibrium configurations}\label{ss:sol}

\begin{figure*}[ht!]
\centering
\includegraphics[width=8.5cm]{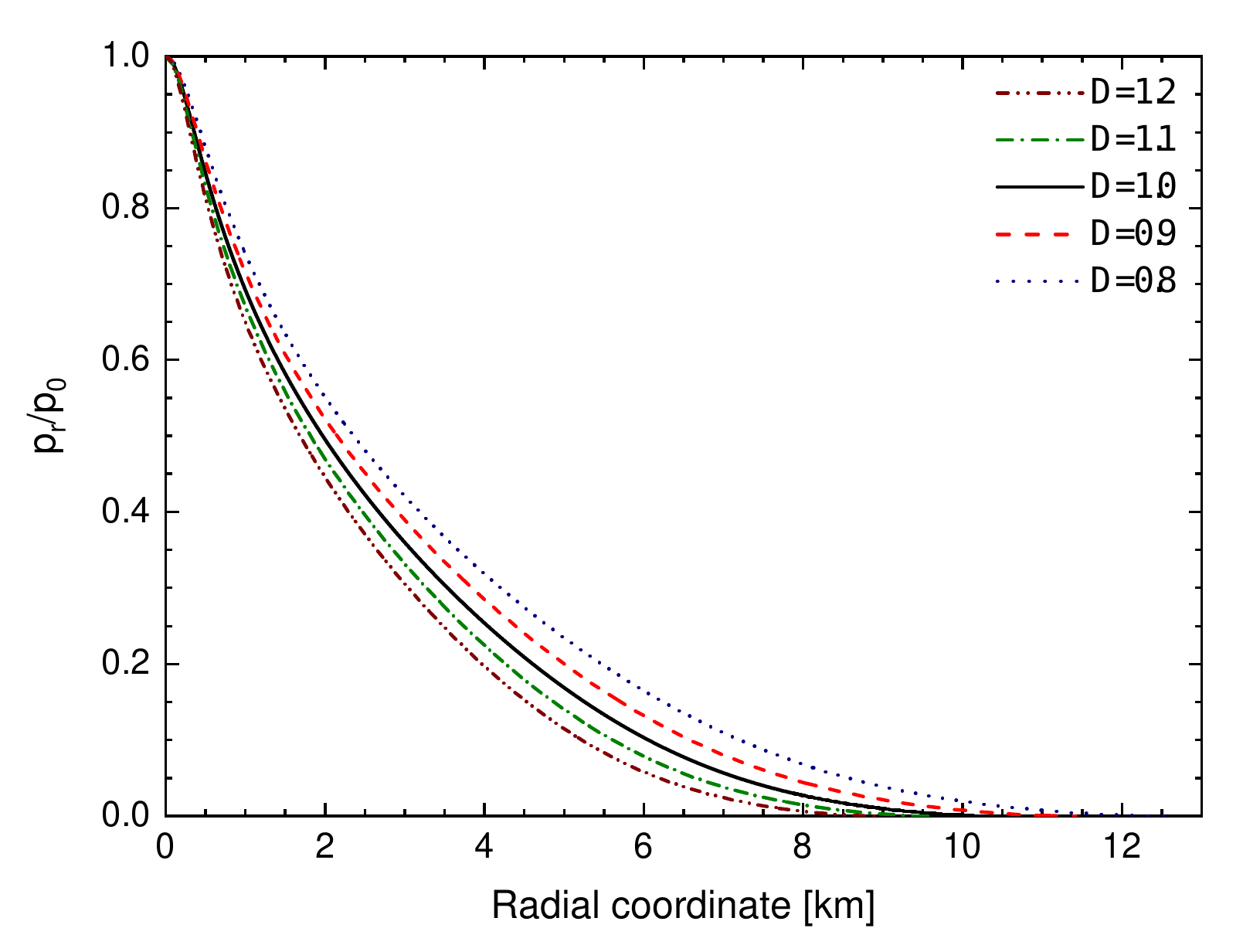} 
\includegraphics[width=8.5cm]{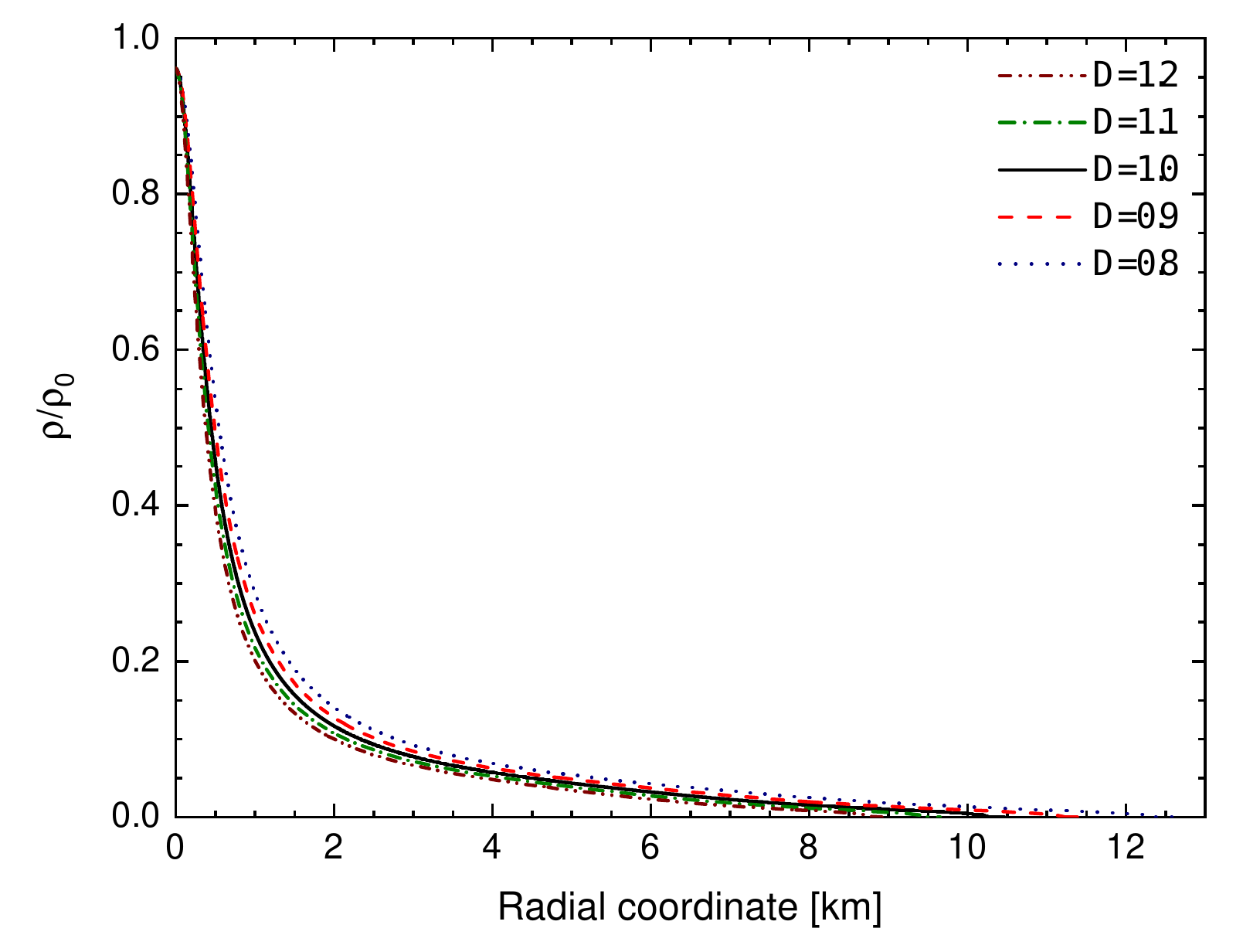}
 \includegraphics[width=8.5cm]{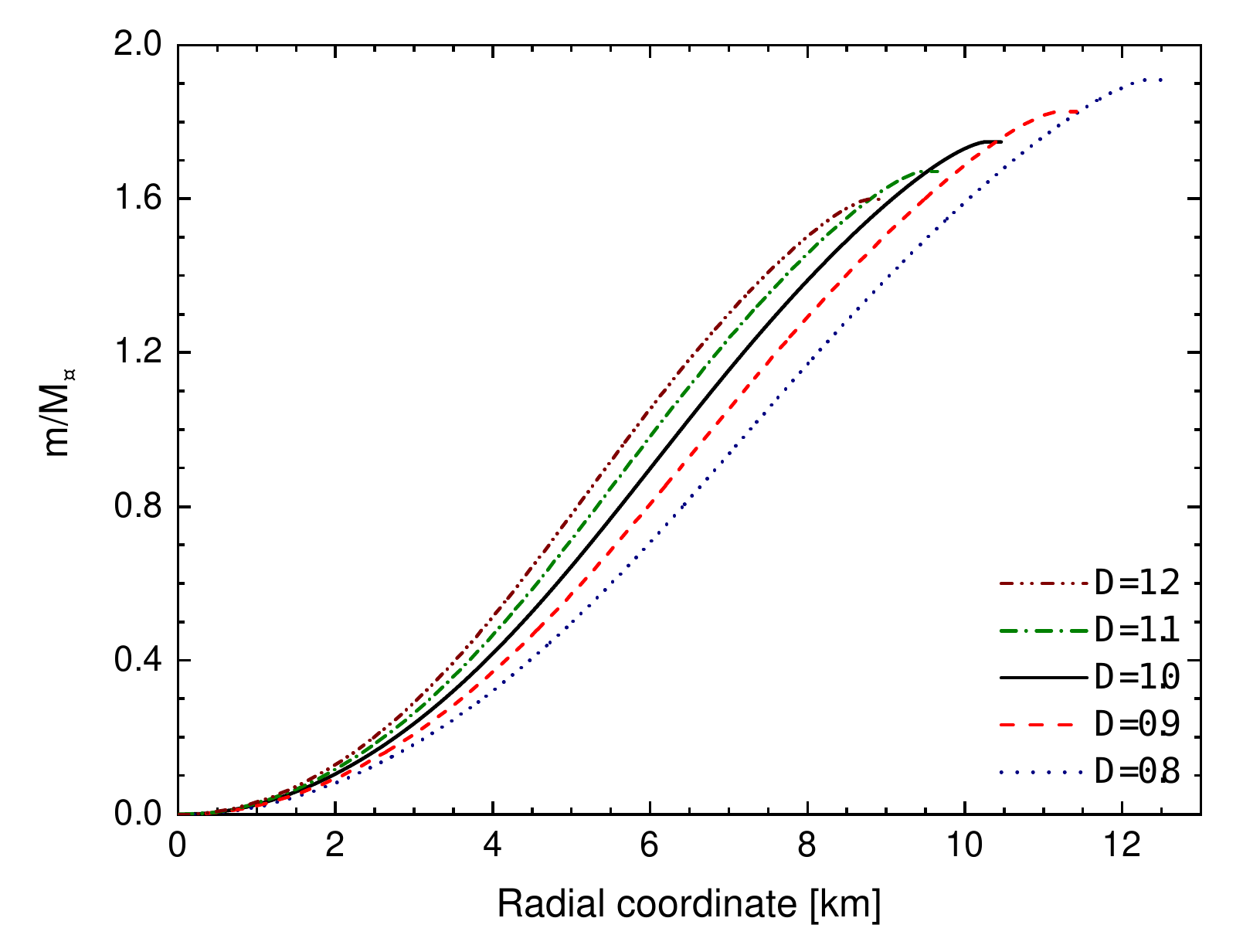} 
\includegraphics[width=8.5cm]{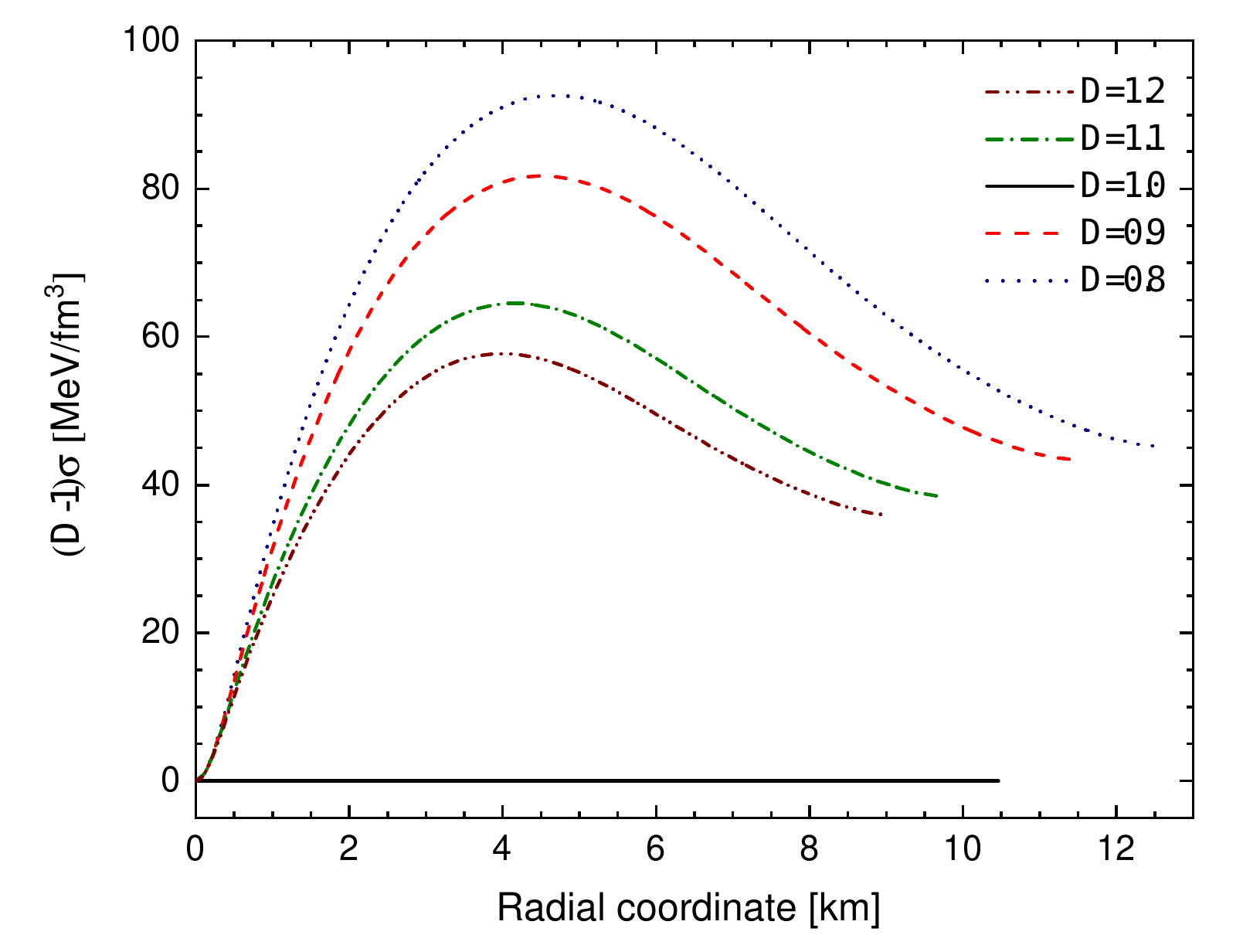} 
\caption{\label{fig1} On the upper panels we present the radial pressure and energy density profiles in their normalized forms against the radial coordinate. The bottom panels show the mass, in solar masses $M_{\odot}$, and the anisotropic factor profiles as a function of the radial coordinate. In all panels, five different values of $\mathcal{D}$ are employed. The pressure and energy density of normalization are $p_0=1000\,[\rm MeV/fm^3]$ and $\rho_0=20000\,[\rm MeV/fm^3]$.}
\end{figure*}

The profiles of the radial pressure $p_r/p_0$, energy density $\rho/\rho_0$, mass $m/M_{\odot}$, and anisotropy $(\mathcal{D}-1)\sigma$ as a function of the radial coordinate are shown in Fig.\ref{fig1} for different values of $\mathcal{D}$. The normalization factor for the pressure and energy density are $p_0=1000\,[\rm MeV/fm^3]$ and $\rho_0=20000\,[\rm MeV/fm^3]$, respectively. On the top panels of Fig. \ref{fig1}, as expected, for $\mathcal{D}=1$, the radial pressure and the energy density decay monotonically with the increment of $r$. For $\mathcal{D}\neq1$, the fluid radial pressure and the energy density slightly change with $\mathcal{D}$. In some intervals of $r$, these two functions grow and diminish their values with the respective increment or diminution of $\mathcal{D}$. 

On the bottom panels of Figure \ref{fig1}, as habitual, for $\mathcal{D}=1$, the mass grows monotonically until attains the surface of the star and the anisotropic profile is null along the star. For $\mathcal{D}\neq1$, $m/M_{\odot}$ rises (declines) with the diminution (raising) of $\mathcal{D}$ and the anisotropic profile is non-null.

\begin{figure}
\centering
\includegraphics[width=8.5cm]{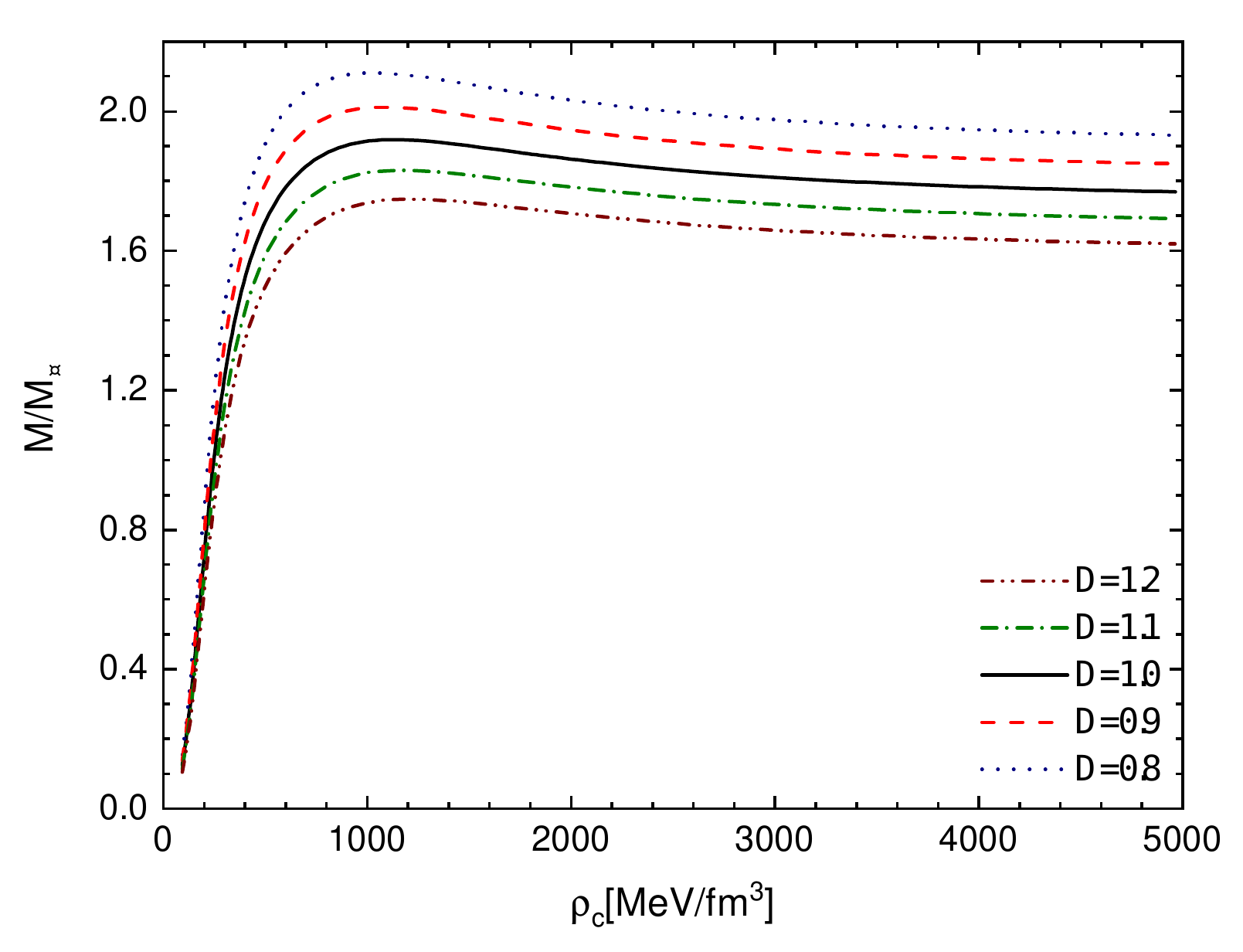} 
\caption{\label{fig_MRho} Total mass $M$, in solar masses $M_{\odot}$, as a function of the central energy density for some different values of $\mathcal{D}$.}
\end{figure}

The total mass, in solar masses $M_{\odot}$, against the central energy density $\rho_c$ is plotted in Figure \ref{fig_MRho} for different values of the parameter $\mathcal{D}$ for $90<\rho_c<5000\,[\rm MeV/fm^3]$. The mass grows monotonically with $\rho_c$ until attains a maximum mass point. After this point, $M/M_{\odot}$ decays with the increment of $\rho_c$. The profile $M(\rho_c)$ changes with $\mathcal{D}$. In the central energy density interval employed, when $\mathcal{D}$ is increased, lower total masses are found. On the other hand, when $\mathcal{D}$ is diminished, larger total masses are determined.

\begin{figure}
\centering
\includegraphics[width=8.5cm]{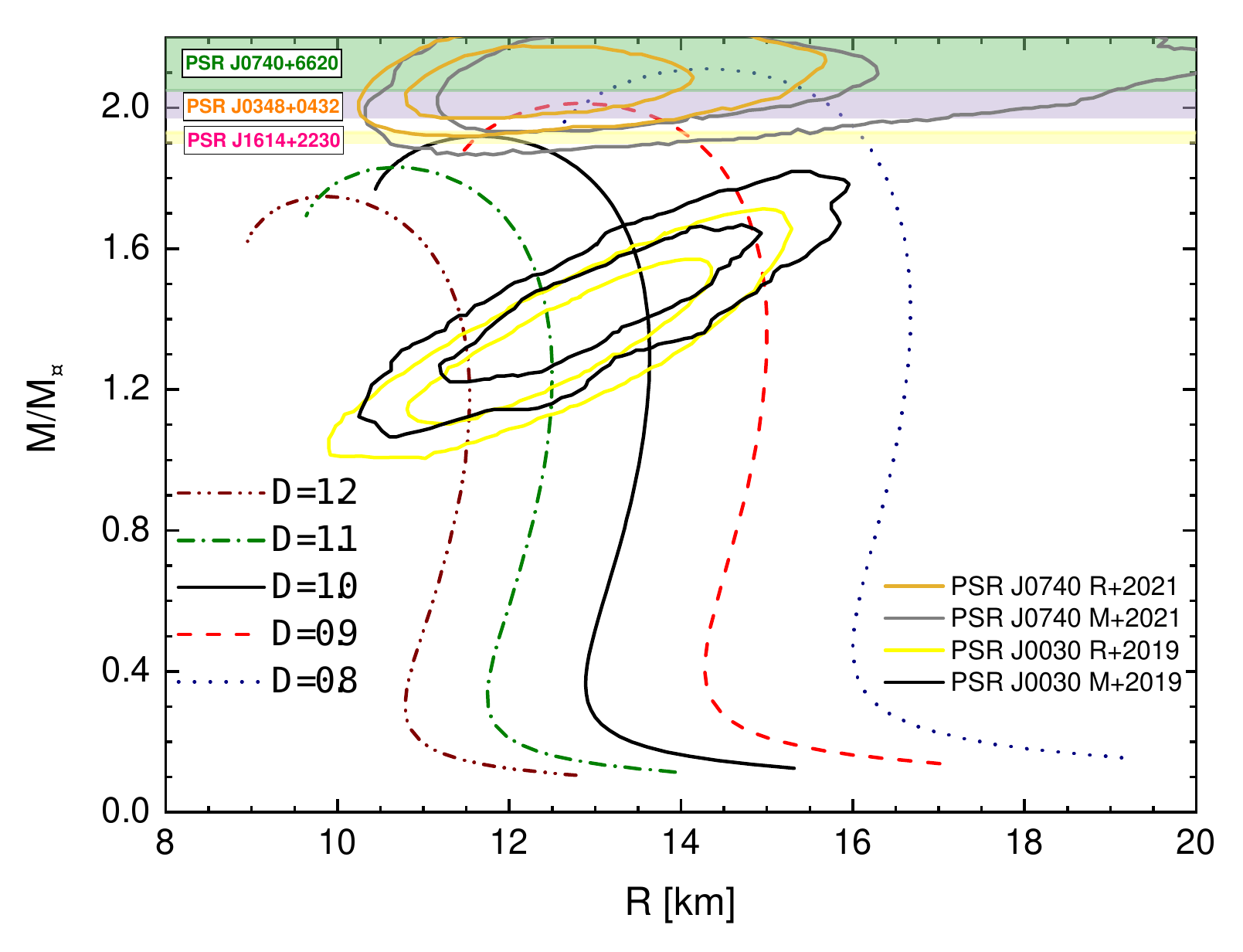} 
\caption{\label{fig_MR} The total stellar mass, in solar masses $M_{\odot}$, versus the total radius for five values of $\mathcal{D}$.}
\end{figure}

In Fig.\ref{fig_MR}, the total mass normalized in solar masses as a function of the radius for some values of $\mathcal{D}$ is presented. The $M(R)$ function begins to grow with the decrease of the total radius until it reaches a minimum total radius. At this point, the curve turns clockwise for $M(R)$ to increase its value with the growth of the total radius until reaching a maximum total radius. After this point, the curve turns counterclockwise, so that $M(R)$ begins to grow with decreasing $R$ until reaching the maximum total mass. From this point onwards, $M(R)$ decreases with the reduction of $R$. For a fixed central energy density, larger (lower) values of $\mathcal{D}$ allow us to find stars with smaller (larger) mass and radius. We also depict the data obtained by NICER for the pulsars PSR J0030 + 0451 \cite{riley/2019} and PSR J0740 + 6620 \cite{riley2, miller/2021}. The image also features bands for the pulsars PSR J0740 + 6620 \cite{cromartie}, PSR J0348 + 0432 \cite{antoniadis}, and PSR J1614 + 2230 \cite{demorest}.

Table \ref{tablei} shows some equilibrium configurations. The constant $\mathcal{D}$, the maximum masses with their respective central energy density $\rho_c$, and equatorial $R$ and polar $Z$ total radii are presented. We note that the total mass and the radii diminish (increase) with the increment (decrement) of $\mathcal{D}$. However, the central energy density becomes greater (smaller) with the growth (decrease) of $\mathcal{D}$.

\begin{table}[h] 
\centering
\begin{tabular}{ccccccccc}
\hline\hline
$\mathcal{D}$  & & $M_{\rm max}/M_{\odot}$ & & $\rho_c\,[\rm MeV/fm^3]$ & & $R\,[\rm km]$& & $Z\,[\rm km]$\\\hline
$1.2$ & & $1.7481$ & & $1202.9$ & & $9.8702$ & & $11.844$\\
$1.1$ & & $1.8306$ & & $1158.7$ & & $10.704$ & & $11.774$\\
$1.0$ & & $1.9188$ & & $1117.5$ & & $11.683$ & & $11.683$\\
$0.9$ & & $2.0123$ & & $1078.9$ & & $12.848$ & & $11.563$\\
$0.8$ & & $2.1103$ & & $1042.6$ & & $14.264$ & & $11.411$\\\hline\hline
\end{tabular}
\caption{\label{tablei} Values employed for the parameter $\mathcal{D}$  and the static equilibrium configuration with maximum masses with their respective central energy density and the equatorial and polar total radii.}
\end{table}



\section{Discussion and concluding remarks}\label{sec:dc}

The robustness of NS hydrostatic equilibrium configuration models increases when taking into account that the density regime in such compact objects may imply in anisotropy inside them \cite{ruderman/1972,canuto/1974} (check also References \cite{heintzmann/1975,hillebrandt/1976,rahmansyah/2020,bordbar/2022}). Anisotropy in NSs has been constrained through gravitational waves \cite{roupas/2021} and multimessenger astrophysics \cite{rahmansyah/2021}. It was shown, in Reference \cite{biswas/2019}, that anisotropic pressure can significantly reduce the tidal deformability of NSs . Recently, in Reference \cite{becerra/2024}, anisotropy inside NSs has been constrained from the calculation of the radial and tangential sound velocities.

Anisotropy can lead to deformation of the star. The next step in constructing trustworthy NSs equilibrium configurations is to consider, therefore, that these objects are deformed. This has been done in the present article.

Our first step was to drop the assumption of spherical symmetry, which is denoted by working out Equation \eqref{dtov1}. Then, from an appropriate choice for the energy-momentum tensor (check Equation \ref{tmomentum}), we have derived what we called ``$\mathcal{D}$-TOV equations'', which are the analogous of the usual TOV equation for deformed compact astrophysical objects. These equations were solved for the GM1 EoS.

In Fig.2, we presented the relation between $r$ and pressure, energy density, mass and anisotropy factor for different values of $\mathcal{D}$. Both pressure and energy density decrease in the direction of increasing $r$. 

We can observe, in Fig.3, that lower values of the deformation parameter $\mathcal{D}$ imply in higher total masses for $\mathcal{D}$NSs. For $\mathcal{D} = 0.8$, for instance, our results show a total mass of 2.1103M$_\odot$ and energy density of 1042.6 MeV/fm³.

In Fig.\ref{fig_MR} we have constructed the mass vs. radius diagram for different values of $\mathcal{D}$. Particularly, the maximum mass expected for $\mathcal{D}$NSs increases with the decreasing of $\mathcal{D}$ and decreases with its increasing.

Figure \ref{fig_MR} also contains observational data of seven different pulsars from which it is possible to constrain the acceptable values of $\mathcal{D}$. From the confrontation of the theoretical predictions of our model with observational data, we notice that values slightly smaller than $0.9$ for the parameter $\mathcal{D}$ are favoured. 

If the secondary object in GW190814 is confirmed as a NS, one can model $\mathcal{D}$NSs for ``stiffer'' E'soS and verify the possibility of attaining the $\sim2.6$ mass scale predicted in the event.

The field of deformed compact objects is emergent. Different approaches have been presented with the purpose of constructing self-consistent non-spherical models for compact stars. Sometimes these models are confronted with observational data.

Zamani and Bigdeli have described deformation by expanding the spherically symmetric metric in multipoles up to the quadrupole term \cite{zamani/2019}. Alvear Terrero et al. used a metric similar to ours (\ref{dtov1}) but applied it for magnetized strange stars \cite{terrero/2021}.  

Here, in the present article, we have assumed deformation is caused by anisotropy, which can appear in regimes of very high density. Further studies may consider the magnetic field contribution, with different E'soS. Then, the polar and equatorial radii stability can also be investigated. \\

\begin{acknowledgments}
JTQ would like to express sincere gratitude to the Coordenação de Aperfeiçoamento de Pessoal de Nível Superior (CAPES) for the financial support, making this research possible. JDVA thanks Universidad Privada del Norte and Universidad Nacional Mayor de San Marcos for the financial support - RR Nº$\,005753$-$2021$-R$/$UNMSM under the project number B$21131781$. The Authors are thankful to J.E. Horvath for numerous discussions regarding deformed neutron stars without which the present paper would not be written.
\end{acknowledgments}


\end{document}